%% file: main.tex
\documentclass[aps,reprint,twocolumn,10pt,superscriptaddress,english,showpacs,longbibliography,nofootinbib,floatfix]{revtex4-1}
\usepackage{amsmath}
\usepackage{amssymb}
\usepackage{amsthm}
\usepackage{amsfonts}
\usepackage{listings}
\usepackage{longtable}
\usepackage{enumerate}
\usepackage{latexsym}
\usepackage{color}
\usepackage{multirow}
\usepackage[figuresright]{rotating}
\usepackage[table]{xcolor}
\usepackage{setspace} 
\usepackage{blindtext}
\usepackage{dcolumn}
\usepackage{booktabs}  
\usepackage{bbding}  
\usepackage{geometry}  
\usepackage{braket}
\usepackage{bbm}
\usepackage{array,etoolbox}
\usepackage{chemformula}
\let\ce\ch
\preto\tabular{\setcounter{magicrownumbers}{0}}
\newcounter{magicrownumbers}

\def\ie{{\it i.e.},\ }

\usepackage{bm}
\usepackage{hyperref}
\hypersetup{
 pdfnewwindow=true, colorlinks=true,
 linkcolor=blue, anchorcolor=blue,
 citecolor=blue, filecolor=blue,
 menucolor=blue, urlcolor=blue}
\definecolor{red}{rgb}{1,0,0}
\definecolor{blue}{rgb}{0,0,1}
\definecolor{black}{rgb}{0,0,0}

\usepackage{graphicx}
\usepackage{subfigure}
\usepackage{longtable}
\usepackage{chemformula}
\usepackage{appendix}
\usepackage{comment}
\usepackage[utf8]{inputenc}
\usepackage{bibunits}
\input{macros}

\begin{document}

\title{Symmetry-guided catalogue of chiral phonon materials}

\author{Yue Yang}
\thanks{These authors contributed equally}
\affiliation{New Cornerstone Science Laboratory, Center for Correlated Matter and School of Physics, Zhejiang University, Hangzhou, China}

\author{Zhenyu Xiao}
\thanks{These authors contributed equally}
\affiliation{International Center for Quantum Materials, School of Physics, Peking University, Beijing, China}
\affiliation{Princeton Quantum Initiative, Princeton University, Princeton, NJ, USA}

\author{Yu Mao}
\thanks{These authors contributed equally}
\affiliation{New Cornerstone Science Laboratory, Center for Correlated Matter and School of Physics, Zhejiang University, Hangzhou, China}

\author{Zhanghuan Li}
\affiliation{Beijing National Laboratory for Condensed Matter Physics, Institute of Physics, Chinese Academy of Sciences, Beijing, China}

\author{Zhenyang Wang}
\affiliation{College of Computer Science and Technology, Zhejiang University, Hangzhou, China}

\author{Tianqi Deng}
\affiliation{State Key Laboratory of Silicon and Advanced Semiconductor Materials, School of Materials Science and
Engineering, Zhejiang University, Hangzhou, China}
\affiliation{Key Laboratory of Power Semiconductor Materials and Devices of Zhejiang Province, Institute of Advanced Semiconductors, ZJU-Hangzhou Global Scientific and Technological Innovation Center, Zhejiang University,
Hangzhou, China}

\author{Yanhao Tang}
\affiliation{School of Physics, Zhejiang University, Hangzhou 310058, China}
\affiliation{Zhejiang Key Laboratory of Micro-Nano Quantum Chips and Quantum Control, Zhejiang University, Hangzhou, China}

\author{Zhi-Da Song}
\affiliation{International Center for Quantum Materials, School of Physics, Peking University, Beijing, China}

\author{Yuan Li}
\affiliation{Beijing National Laboratory for Condensed Matter Physics, Institute of Physics, Chinese Academy of Sciences, Beijing, China}

\author{Huiqiu Yuan}
\affiliation{New Cornerstone Science Laboratory, Center for Correlated Matter and School of Physics, Zhejiang University, Hangzhou, China}
\affiliation{Institute of Fundamental and Transdisciplinary Research, Zhejiang University, Hangzhou, China}
\affiliation{Institute for Advanced Study in Physics,Zhejiang University, Hangzhou, China}
\affiliation{State Key Laboratory of Silicon and Advanced Semiconductor Materials, Zhejiang University, Hangzhou, China}

\author{Ming Shi}
\affiliation{New Cornerstone Science Laboratory, Center for Correlated Matter and School of Physics, Zhejiang University, Hangzhou, China}

\author{Yuanfeng Xu}
\email{y.xu@zju.edu.cn}
\affiliation{New Cornerstone Science Laboratory, Center for Correlated Matter and School of Physics, Zhejiang University, Hangzhou, China}

\begin{abstract}
Chiral phonons—circularly polarized lattice vibrations with angular momentum—have become a key frontier in quantum materials. They offer ways to control heat and information through their coupling to electronic spin, orbital, and valley degrees of freedom. We present a symmetry-based framework that classifies phonon chirality across all crystallographic space groups. Our approach, built on symmetry representations of phonon angular momentum in reciprocal space, identifies three phononic material classes, namely, crystals with no chirality, chiral crystals with conventional $s$-wave helicity, and a group of achiral crystals hosting exotic higher-order helicities such as $d$, $g$ and $i$ waves. Through high-throughput computation, we shortlist the most promising material candidates for experimental investigation. These results are compiled into the open-access Chiral Phonon Materials Database, which enables screening for materials with the desired chiral phonon properties. Our work establishes both the theoretical framework and the material platform required to exploit the properties of chiral phonons for next-generation thermal management and quantum technologies.
\end{abstract}

\maketitle

\section{Introduction}

Chirality, a fundamental property that prevents an object from superimposing on its mirror image\cite{kelvin1894molecular}, pervades nature, from the double helix of DNA\cite{watson1953molecular} to the stereoselectivity in asymmetric catalysis\cite{knowles1968catalytic,horner1968asymmetric} and the topologically protected edge states in quantum Hall systems\cite{RevModPhys.58.519}. In recent years, the chirality of phonons has emerged as a frontier in condensed-matter physics due to its profound implications for spintronics, valleytronics, thermal management and phononic engineering. Chiral phonons\cite{zhang2014angular,zhang2015chiral,zhu2018observation,ishito2023truly,zhang2025new}, analogous to circularly polarized electromagnetic waves, manifest themselves as circularly polarized vibrational modes carrying intrinsic angular momentum. In crystalline materials, a chiral phonon with finite lattice momentum is characterized by its helicity $h(\textbf{q}):=\textbf{e}_q\cdot\textbf{J}^{ph}(\textbf{q})$, the scalar product of the unit vector along the lattice momentum ($\textbf{e}_{q}$) and phonon angular momentum ($\textbf{J}^{ph}(\textbf{q})$). 
A phonon exhibits positive helicity if its angular momentum aligns with the momentum direction, and negative helicity if they are anti-aligned.
This chirality not only governs the dynamic coupling between phonons and other quasiparticles (for example, electron, photon, or magnon) but also dictates selection rules in scattering processes, potentially enabling remarkable phenomena such as the Einstein-de Haas effect\cite{tauchert2022polarized,dornes2019ultrafast,choi2024real,zhang2025measurement,chen2022chiral,hamada2018phonon,PhysRevB.106.115102}, spin Seebeck effect\cite{uchida2008observation,adachi2013theory,kim2023chiral,fransson2023chiral,li2024chiral,nishimura2025theory}, chiral-induced spin selectivity effect \cite{naaman2012chiral,naaman2019chiral,liu2021chirality,evers2022theory}, thermal Hall effect\cite{strohm2005phenomenological,kagan2008anomalous,grissonnanche2020chiral}, and phononic magnetism\cite{xiong2022effective,long2018intrinsic,luo2023large,PhysRevB.107.L020406,PhysRevResearch.4.013129,PhysRevLett.128.075901,hernandez2023observation,yao2025theory,PhysRevB.110.094401,chaudhary2026anomalous,mustafa2025origin}.

Chirality and symmetry are intrinsically interrelated properties in physical systems.
In crystalline solids, the geometric chirality of a crystal is determined by its space group(SG): a crystal is chiral if its SG is a Sohncke group (that is, one that lacks any improper rotation operations such as inversion or roto-inversion), whereas an achiral crystal possesses at least one improper rotation operation.
Within three-dimensional space, the crystalline structures are classified into 65 Sohncke groups and 165 achiral (non-Sohncke) groups. In an achiral crystal, the atomic displacements associated with a phonon mode exhibiting a non-trivial improper-rotation eigenvalue (that is, an eigenvalue distinct from that corresponding to the identity) can locally break the achiral symmetry, inducing a transient chiral distortion\cite{romao2024phonon}. This symmetry-breaking can be initiated through interactions with photons, magnons, or comparable excitations. By contrast, in a chiral crystal, every phonon mode is inherently chiral, as its dynamic lattice vibrations typically further reduce, or at minimum maintain, the already diminished symmetry present in the structure.

Despite a decade of theoretical predictions and experimental investigations of chiral phonon materials\cite{zhang2015chiral,ishito2023truly,SiO2,zhu2018observation,hamada2018phonon,du2019lattice,zhang2023weyl,chen2021propagating,tauchert2022polarized,PhysRevLett.132.056302,PhysRevB.105.104301,che2025magnetic,Pols2024acs.jpclett.4c01629,Pols2025acs.nanolett.5c01708,PhysRevMaterials.6.104410,ueda2025chiral}, explicitly characterized candidates remain comparatively scarce compared to the vast repository of crystalline materials catalogued in structural databases. Expanding this material base is, therefore, essential to fully realize the potential of chiral phonons in fundamental research and practical applications. Consequently, there exists an urgent need to establish a comprehensive classification methodology for chiral phonon materials and to conduct systematic, high-throughput computational searches to identify promising candidates.

In the present work, we address these challenges by developing a rigorous symmetry-based classification of chiral phonon materials that encompass all 230 SGs, elucidating the precise group-theoretical constraints that govern the emergence and characteristics of chiral phonons. We subsequently implement a high-throughput computational workflow — integrating precomputed and newly generated \textit{ab initio} phonon data — to systematically catalogue chiral phonon materials across \nbrofmats\ crystalline compounds. Our results not only provide a universal theoretical framework for classifying chiral-phonon phenomena but also offer an accessible database of candidate materials with detailed phonon-chirality characterizations, paving the way for future experimental validation and applications.

\section{Classification methods}
\label{sec: classify method}

\begin{figure*}[ht]
    \centering
    \includegraphics[width=0.9\linewidth]{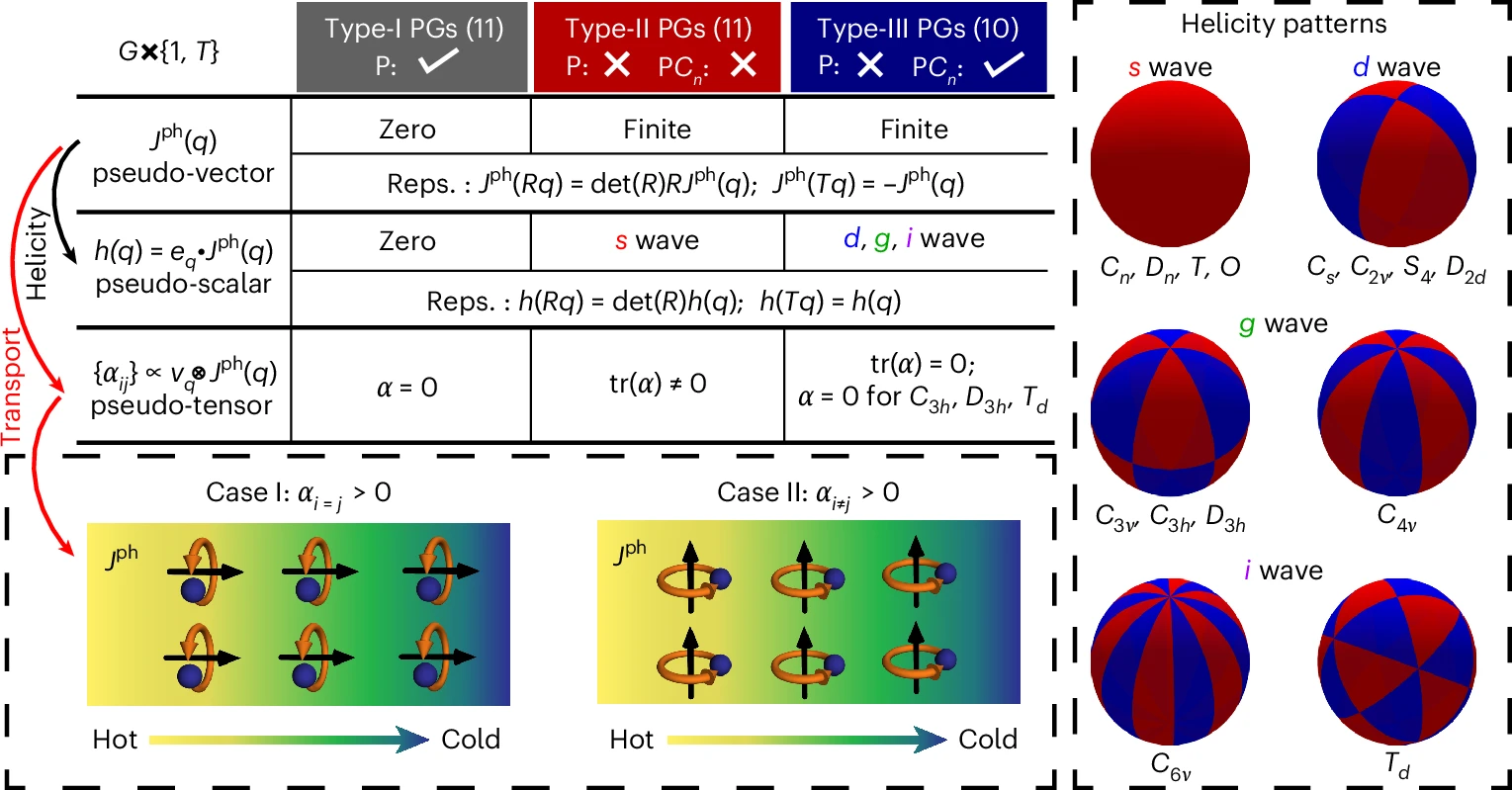}
    \caption{\textbf{Summary of the classifications.} The 32 PGs ($G$) of the 230 SGs are categorized into three types:: 11 centrosymmetric PGs (type I) with inversion ($\mathcal{P}$), 11 chiral PGs (type II) lacking $\mathcal{P}$ or any roto-inversion ($\mathcal{P}C_n$), and 10 non-centrosymmetric achiral PGs (type III) without $\mathcal{P}$ but with $\mathcal{P}C_n$). For each type, we derive symmetry constraints on the phonon angular momentum $\mathbf{J}^{ph}$, helicity ($h$), and the velocity-angular momentum tensor ($\mathbf{v}\bigotimes\mathbf{J}^{ph}$) that determines the phonon magnetization response in thermal transport ($\alpha$), all based on their transformation properties under PG operations ($\mathcal{R}$) and time-reversal symmetry ($\mathcal{T}$). Right panel: Helicity patterns around the Brillouin zone center for type-II and type-III PGs. Red and blue regions signify helicity of opposite sign. The polynomials that characterizing these helicity patterns are tabulated in Supplementary Information \cite{SI}, section A. Bottom panel: Schematics of two symmetry‐allowed chiral‐phonon transport responses. $\alpha_{i=j}>0$ ($\alpha_{i\neq j}>0$) carries phonon angular momentum parallel (perpendicular) to the thermal gradient.}
    \label{fig:helicity}
\end{figure*}

In this section, we systematically classify phononic chirality across all 230 SGs by analysing the symmetry representations of phonon angular momentum, which characterize circularly polarized lattice vibrations in reciprocal space. We subsequently derive the symmetry representations of phonon helicity, relevant to Raman scattering experiments, as well as the velocity–angular-momentum tensor that underlies phonon magnetism in thermal transport. The main results are summarized in Fig. \ref{fig:helicity}, with detailed discussions below.

We first investigate the symmetry properties of the phonon angular momentum $\textbf{J}^{ph}(\textbf{q})$ in the momentum space.
Under inversion symmetry ($\mathcal{P}$), $\textbf{q}$ reverses its direction, whereas $\textbf{J}^{ph}$ is invariant. By contrast, under time-reversal symmetry (TRS; $\mathcal{T}$), both $\textbf{q}$ and $\textbf{J}^{ph}$ reverse directions.
Therefore, in non-magnetic materials with inversion symmetry, the operation $\mathcal{PT}$ requires $\textbf{J}^{ph}(\textbf{q}) = -\textbf{J}^{ph}(\textbf{q})$; thus $\textbf{J}^{ph}(\textbf{q})$ must vanish for any $\textbf{q}$ \cite{PhysRevB.108.134307}.
Let $G$ be the PG (PG) of the SG $\mathcal{G}$ without inversion $\mathcal{P}$.
In such groups, $\textbf{J}^{ph}(\textbf{q})$ is generally non-zero for a generic $\textbf{q}$. A proper (improper) spatial rotation $\mathcal{R}$ with $\det(\mathcal{R}) = 1$ ($-1$) in $G$ requires that 
\begin{equation}
    \textbf{J}^{ph}(\mathcal{R} \textbf{q}) = \det(\mathcal{R}) \mathcal{R} \textbf{J}^{ph}(\textbf{q}) .
\end{equation}
In addition, TRS requires that $\textbf{J}^{ph}(\textbf{q}) = -\textbf{J}^{ph}(-\textbf{q})$. 
Thus, the PG $G$ of an SG does not completely capture the symmetry of $\textbf{J}^{ph}(\textbf{q})$ in momentum space. 
The symmetry should be characterized by $G \times C_i$, where the inversion group $C_i$ in the momentum space is generated by TRS. 

For a given $\textbf{q}$, the operations in $G \times C_i$ that remain momentum $\textbf{q}$ invariant up to a translation by a reciprocal lattice vector constitute its little PG $G_{\textbf{q}}$. In particular, $G_{\textbf{q}}$ should be isomorphic to the little magnetic group of the associated grey group (that is, type-II magnetic SG) of the SG $\mathcal{G}$ because of additional TRS.
The angular momentum $\textbf{J}^{ph}(\textbf{q})$ at $\textbf{q}$ forms a three-dimensional real representation $\gamma_{G_\textbf{q}}$ of $G_{\textbf{q}}$, which can be decomposed to a direct sum of irreducible representations. 
Supplementary Information \cite{SI}, section A lists the decompositions. If trivial irreducible representations (that is, identity representations) appear in the decomposition, the corresponding components of $\mathbf{J}^{ph}$ can be non-zero.
Conversely, components associated with non-trivial irreducible representations must vanish. 

Next, we investigate the symmetry of helicity $h$, defined as $h(\textbf{q})=\textbf{e}_q\cdot\textbf{J}^{ph}(\textbf{q})$, with $ \textbf{e}_q$ being the unit vector along the wave vector $\textbf{q}$.
As a product between a polar vector and a pseudovector, $h$ is a pseudoscalar: $h(\mathcal{R} \textbf{q}) = \det(\mathcal{R} ) h(\textbf{q})$ for a rotation $\mathcal{R}$ in $G$. 
In addition, $h$ is invariant under TRS: $ h(\textbf{q}) = h (-\textbf{q})$. 
Similar to $\textbf{J}^{ph}$, if inversion exists in $G$ (referred to as type-I PG), $\mathcal{PT}$ symmetry requires $ h(\textbf{q}) = 0$ for any $\textbf{q}$.
If there is no improper rotation or inversion in $G$ (referred to as type-II PG), $h(\textbf{q})$ is invariant under all operations in $G$, as well as TRS: $h(\textbf{q})$ forms an $s$-wave-type pattern in the Brillouin zone.
By contrast, if improper rotation exists but inversion is absent (referred to as type-III PG), $h(\textbf{q})$ should form an alternating pattern. The pattern can be like a $d$, $g$ or $i$ wave (Fig.~\ref{fig:helicity}). In addition, we introduce a distinct transport-related quantity, denoted as $h^{\prime} = \textbf{v}\cdot\textbf{J}^{ph}$, where $\textbf{v} = \nabla \omega(\textbf{q})$ is the phonon group velocity. This quantity characterizes the transport-oriented chiral properties; because both the lattice momentum and the group velocity are polar vectors, $h^{\prime}$ and $h$ share identical symmetry properties in the Brillouin zone. Note that $h^{\prime}$ can vanish at momenta at which the group velocity is zero. 

Finally, we consider the symmetry of the pseudotensor $\textbf{v} \otimes\textbf{J}^{ph}$, which governs the temperature‐gradient‐induced phonon magnetization $\textbf{M}_i = \alpha_{ij} \partial_j T$, where $\partial_j T$ is the temperature gradient along the $j$ direction and $ \textbf{M}_i$ is magnetization along the $i$ direction. 
Within Boltzmann transport theory, the response tensor $\mathbf{\alpha}_{ij}$ is proportional to $ \sum_p \int  \textbf{J}^{ph}_i \textbf{v}_j \partial f_B(\omega_p(\mathbf{q})/T)/\partial T d \textbf{q}$, with $p$ being the phonon band index and $f_B(\cdot)$ being the Bose-Einstein distribution~\cite{hamada2018phonon,PhysRevB.106.115102}. The nine components of $\alpha_{ij}$ furnish the same symmetry representation as the tensor $\mathbf{v}\otimes\mathbf{J}^{\mathrm{ph}}$ at $\mathbf{q}=0$. Any trivial irrep in its decomposition, therefore, corresponds to a symmetry‐allowed (potentially non-zero) component or a linear combination of $\alpha_{ij}$.
In all type-I PGs, $\mathcal{PT}$ symmetry forces $\alpha_{ij}=0$. 
Among the 21 type-II and type-III PGs, 18 admit at least one trivial irrep and, thus, support non-zero components of $\alpha_{ij}$. 
The three exceptions are $C_{3h}$, $D_{3h}$, and $T_d$, which lack any trivial irrep. 
Supplementary Information \cite{SI}, section A provides the full decomposition of the representation of $\mathbf{v}\otimes\mathbf{J}^{\mathrm{ph}}$ and examples of  $\alpha$ tensor calculation of four representative materials. As schematically shown in Fig.~\ref{fig:helicity}, when a diagonal element is non-zero (case I), the phonon magnetization is parallel to the thermal-gradient direction, whereas when an off-diagonal element is non-zero (case II), the phonon magnetization is perpendicular to the thermal-gradient direction. Moreover, in type-II groups the trace $h^{\prime} = {\rm Tr}(\textbf{v} \otimes\textbf{J}^{ph})$ is itself a trivial irrep, guaranteeing at least one non-zero diagonal entry of $\alpha$. 
By contrast, for type-III groups, $h'$ transforms as a non-trivial irrep, so all diagonal components of $\alpha$ vanish in the principal‐axis frame.

It is worth noting that cycloidal phonons\cite{juraschek2025chiral,chaudhary2026anomalous}, whose angular momentum is perpendicular to the lattice momentum, also play an essential role in thermal transport responses\cite{hamada2018phonon}.
Unlike helical phonons, the distribution of cycloidal phonons in the Brillouin zone is characterized by the cross-product of $\textbf{e}_{q}$ and  $\textbf{J}^{ph}(\textbf{q})$, that is, a polar vector $\textbf{P}=\textbf{e}_{q}\times\textbf{J}^{ph}(\textbf{q})$. Consequently, its symmetry representations are identical to those of the lattice momentum $\textbf{q}$. Supplementary Information \cite{SI}, section A provides details on the symmetry representations of cycloidal phonons.

\section{High-throughput calculations}
\subsection{Workflow}

\begin{figure*}[ht]
    \centering
    \includegraphics[width=0.75\linewidth]{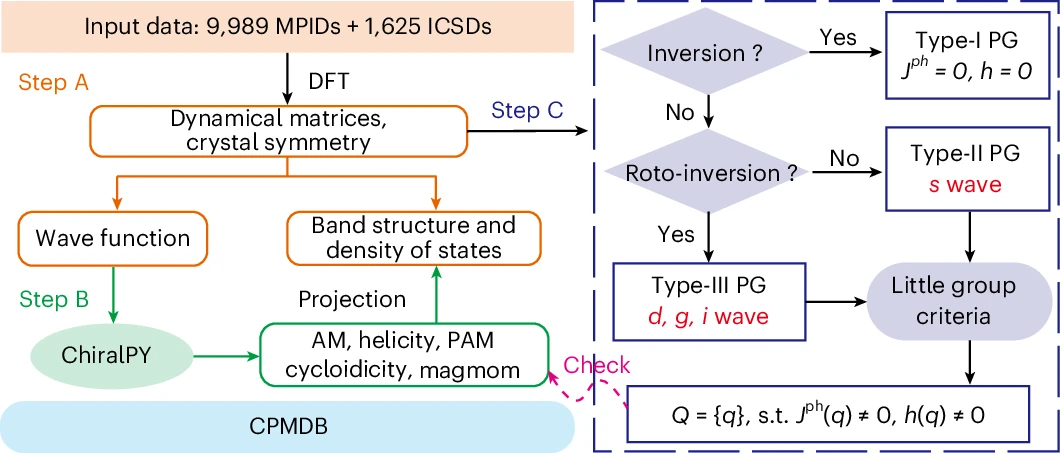}
    \caption{\textbf{Workflow of high-throughput calculations and classifications.} Dynamical matrices for \nbrofmpid\ MPID and \nbroficsd\ ICSD material entries are obtained from the density functional theory (DFT) calculations (Methods). Step A: Crystal symmetry, phonon band structures, density of states and phonon wave functions are analyzed. Step B: Phonon angular momentum (AM), helicity, cycloidicity, PAM, and phonon magnetic moment (magmom) are calculated using the in-house code ChiralPY. Step C: Materials are classified into three categories based on the symmetry analysis from step A—achiral phonons with zero angular momentum (with inversion), chiral phonons with $s$-wave helicity patterns (without inversion or roto-inversion), and chiral phonons with $d$-, $g$-, or $i$-wave helicity patterns (without inversion but with roto-inversion). All the materials' phonon data were collected to build the CPMDB.}
    \label{fig:enter-label}
\end{figure*}

Using the classification methods outlined above, we performed a high-throughput screening, combined with \textit{ab initio} calculations (Methods), to discover chiral phonon materials. The workflow comprises three main steps (Fig.\ref{fig:enter-label}).

In step A, we assembled two input sets: \nbrofmpid\ materials (MPIDs) from the Topological Phonon Database 
(TPDB; \url{https://www.topologicalquantumchemistry.fr/topophonons/index.html})\cite{xu2024catalog} and \nbroficsd\ ICSD entries from the Topological Materials Database (TMDB; \url{https://www.topologicalquantumchemistry.fr/\#/})\cite{vergniory2019complete}. For materials in the TPDB, the force constants were precomputed and archived in the Phonondb data on MDR at NIMS (\url{https://github.com/atztogo/phonondb}), obviating new \textit{ab initio} work. By contrast, although TMDB catalogues tens of thousands of compounds by electronic topology, it lacks phonon data. We, therefore, selected \nbroficsd\ material entries with fewer than ten atoms per unit cell and performed high-throughput \textit{ab initio} calculations to obtain their force constant matrices. For each compound, we extracted the dynamical matrix, the phonon band structure along high-symmetry paths in the Brillouin zone, the phonon density of states, and the corresponding vibrational eigenmodes (that is, phonon wave functions).

In step B, we developed an in-house code named ChiralPY to compute the phonon angular momentum $\textbf{J}^{ph}=(J^{ph}_x,J^{ph}_y,J^{ph}_z)$, the helicity $h$, the cycloidicity vector $\textbf{P}=\textbf{e}_{q}\times\textbf{J}^{ph}(\textbf{q})$ that characterizes cycloidal phonons, pseudo angular momentum (PAM) \cite{zhang2015chiral} and phonon magnetic moment (Methods). PAM is instrumental in characterizing the phase factor that a wave function acquires under (screw) rotations and is crucial in electron-photon-phonon interactions, where both energy and angular momentum must be conserved \cite{zhang2015chiral}. In ionic materials, the magnetic moment induced by chiral phonon modes—via their phonon angular momentum and the gyromagnetic ratio \cite{PhysRevMaterials.3.064405}—provides important information for characterizing phonon magnetization. Feeding the vibrational eigenmodes from Step A into ChiralPY, we then projected $\textbf{J}^{ph}$, $h$, $\textbf{P}$, PAM and magmom onto the phonon band structures.

In step C, we perform a symmetry analysis on each material and classified its phonon band structure into three categories based on the methods in Sec.~\ref{sec: classify method}. 
Materials with inversion symmetry are described by type-I PGs. In such materials, $\mathcal{PT}$ symmetry enforces zero phonon angular momentum at any momenta, rendering all phonons achiral. If the crystals lack both inversion and roto-inversion symmetry, they fall under a type-II PG, generally allowing non-zero phonon angular momentum at generic momenta—except at the eight time reversal invariant momenta points—although the phonon helicity pattern on the isoenergy surface displays an $s$-wave–like character. 
The remaining materials lack inversion symmetry but include roto-inversion symmetry, belonging to type-III PGs. Their phonon band structure can be further distinguished as having $d$-, $g$-, or $i$-wave–like helicity pattern. 
In addition, we cross-checked these symmetry-based classifications against the angular-momentum, helicity, and PAM obtained numerically in Step B to verify consistency with the enforced symmetry constraints.

\subsection{Results and statistics}

Using the classification method and high-throughput computational framework above, we performed a comprehensive survey of phonon chirality across \nbrofmats\ phonon datasets sourced from TPDB (\nbrofmpid\ MPID entries) and TMDB (\nbroficsd\ ICSD entries). This extensive dataset enables a statistically robust and systematic 
mapping of chiral phonon activity throughout the landscape of crystal materials. The results yield several notable and foundational insights.

\begin{figure*}[ht]
    \centering
    \includegraphics[width=0.55\linewidth]{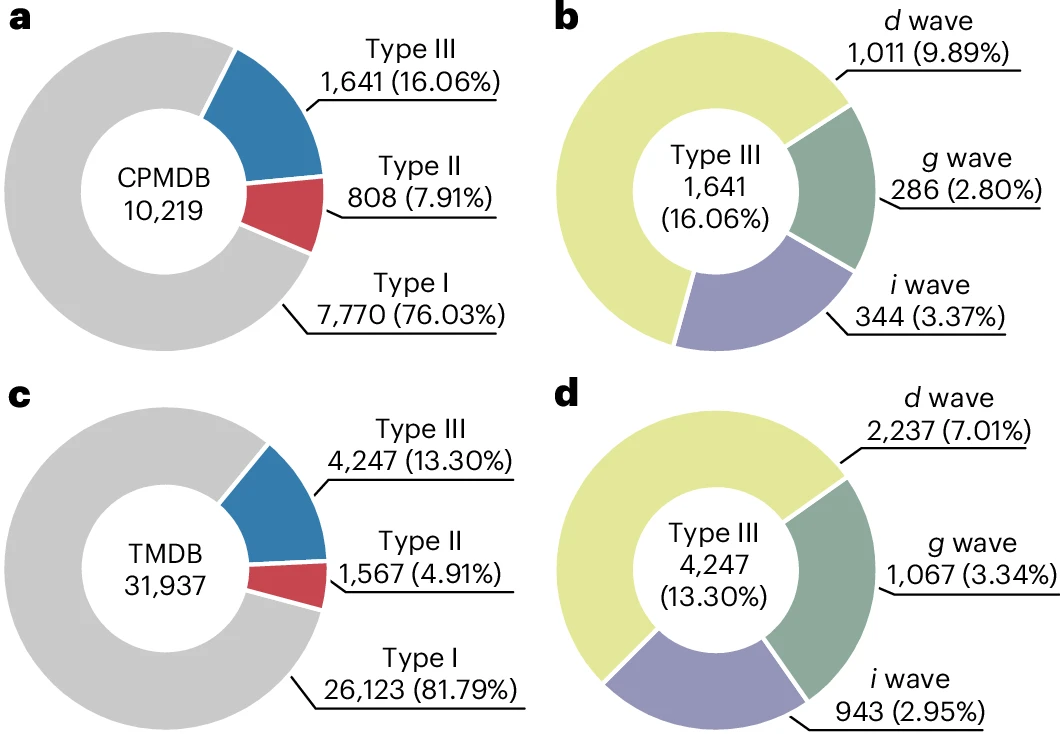}
    \caption{\textbf{Statistical overview of chiral phonon materials in the high-throughput classification. a}, Pie chart showing the percentage distribution of all \nbrofmats\ surveyed materials in the CPMDB across the three phonon chirality types: type I (achiral, \nbroftypeIpercent), type II ($s$-wave, \nbroftypeIIpercent), and type III (\nbroftypeIIIpercent). \textbf{b}, Breakdown of the type-III subset into $d$-wave, $g$-wave and $i$-wave helicity patterns, with corresponding material counts and percentages. \textbf{c},\textbf{d}, Similar statistics as in \textbf{a}(\textbf{c}) and \textbf{b}(\textbf{d}), but for the \nbrofmatstmdb\ materials in TMDB.
    }
    \label{fig3}
\end{figure*}

As shown in Fig. \ref{fig3}a,b, our analysis reveals that \nbroftypeI\ materials (\nbroftypeIpercent) exhibit type-I (achiral) phonon spectra with zero phonon angular momentum and helicity at all wavevectors. 
In sharp contrast, \nbroftypeII\ materials (\nbroftypeIIpercent) are hosted in type-II PGs, which possess only proper rotational symmetries.
They display phonon bands that robustly support non-zero angular momentum and $s$-wave helicity patterns throughout much of the Brillouin zone, except at the eight TRIM points. 
An additional \nbroftypeIII\ materials (\nbroftypeIIIpercent) emerge as type-III candidates, combining the absence of inversion with the presence of at least one improper rotation. 
Notably, these support more intricate chiral phonon textures, giving rise to higher-harmonic helicity patterns—classified here as \nbrofdwave\ (\nbrofdwavepercent) $d$ wave, \nbrofgwave\ (\nbrofgwavepercent) $g$ wave, and \nbrofiwave\ (\nbrofiwavepercent) $i$ wave according to the underlying PG symmetry. Supplementary Information \cite{SI}, section B provides a complete list of chiral phonon materials in type-II and type-III PGs.
The relative scarcity of type‑II and type-III compared to type‑I phonon materials quantitatively reflects the strong crystallographic constraints imposed by the ubiquitous presence of inversion symmetry in crystalline materials.
Our finding also highlights how even subtle deviations from inversion symmetry can unlock entirely new vibrational physics. Importantly, the systematic enumeration establishes a lower bound on the prevalence of chiral phonons in known solid-state materials, underscoring a rich but previously underexplored realm for future research.

Despite differences in the two source databases (TPDB and TMDB), both exhibit qualitatively similar statistical trends across the three phonon chirality classes and within type-III subtypes (see Fig. \ref{fig3}c,d). 
In the TMDB entries, where full phonon spectra are not always available, our symmetry-based analysis nevertheless predicts similar proportions of achiral, $s-$wave, $d-$wave, $g-$wave and $i-$wave chiral phonon materials. This invariance confirms that the encoded phonon chirality is a structural property fundamentally dictated by space-group symmetry—inviting further studies to generalize these predictions to other crystal repositories and synthetic targets.

To maximize the utility of this resource, all computed data—including phonon dispersions, mode-resolved chiral phonon properties, symmetry classifications, and crystallographic information—have been made publicly accessible via the Chiral Phonon Materials Database (CPMDB; \url{https://materialsfingerprint.com/}) developed for this work. Supplementary Information \cite{SI}, section C provides details about the CPMDB. The database’s property-driven search and filtering capabilities empower the community to identify materials matching the desired symmetry, chemical or phononic criteria, accelerating targeted synthesis, characterization, and application efforts.
We further highlight \nbrofideal\ compounds exhibiting pronounced chiral phonon bands and \nbrofmatsmagmom\ compounds possessing large phonon magnetic moments, which are promising candidates for future experimental detection of chiral phonons. The detailed screening criteria and comprehensive materials lists are provided in Supplementary Information \cite{SI}, section D.

\section{Prototype material examples}

\begin{figure*}[ht]
    \centering
    \includegraphics[width=1.0\linewidth]{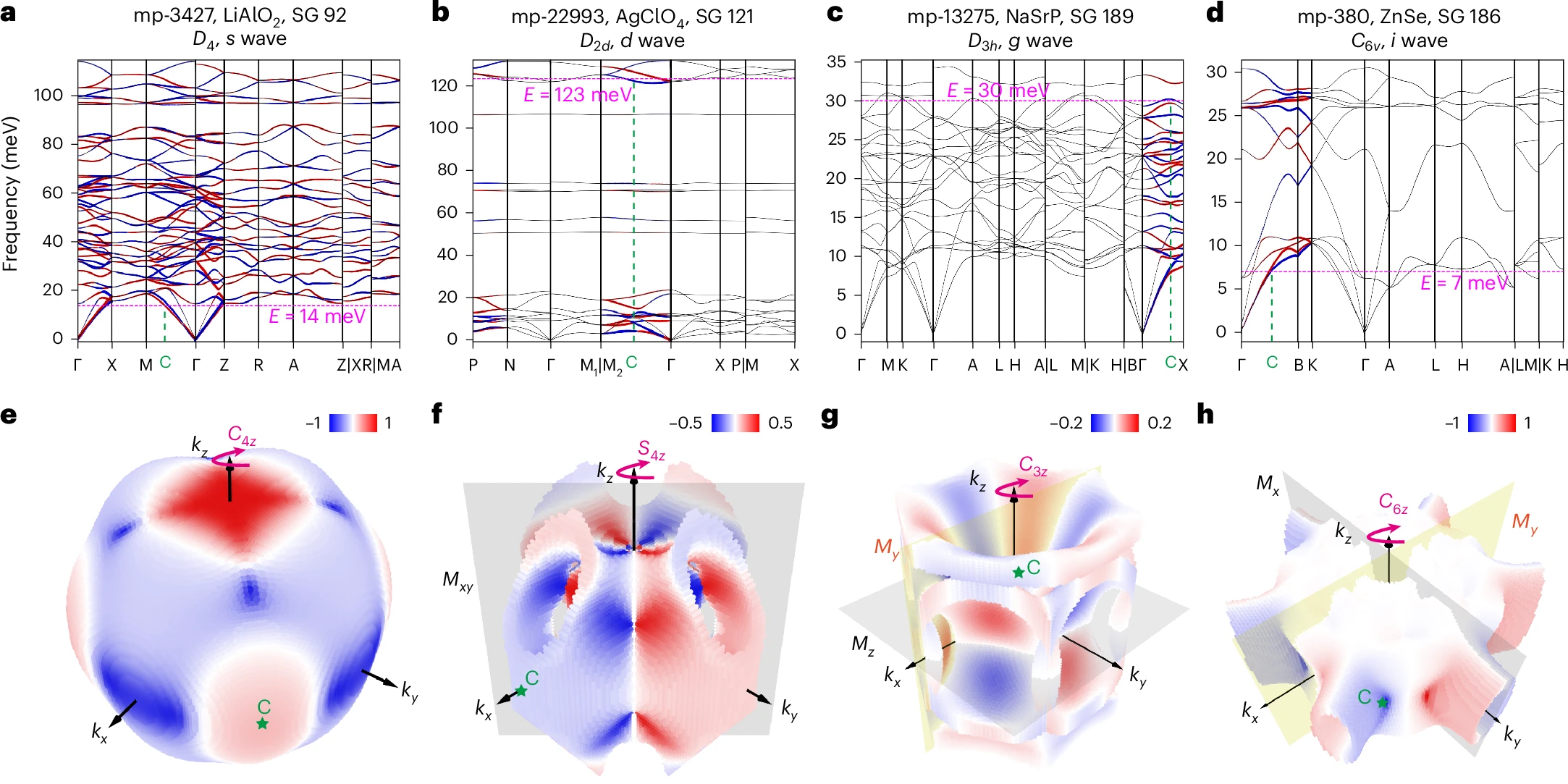}
    \caption{\textbf{Phonon band structures and isoenergy surface 
    with helicity projections for four prototypical materials. 
    a}, \ce{LiAlO2} crystallizes in the tetragonal chiral SG 92 (type-II PG $D_4$), which contains only proper rotations and lacks both inversion and roto-inversion symmetry. 
    This results in an $s$-wave helicity pattern throughout the Brillouin zone, as shown in \textbf{e}. 
    \textbf{b}, \ce{AgClO4} adopts SG 121 (type-III PG $D_{2d}$), which includes roto-inversion symmetries $S_{4z}$ and $M_{xy}$ but no inversion, leading to a $d$-wave helicity pattern characterized by alternating sign regions enforced by symmetry, seen in the helicity-projected isoenergy surface (\textbf{f}). 
    \textbf{c}, \ce{NaSrP}, adopting SG 189, exhibits a type-III PG ($D_{3h}$) that support a $g$-wave helicity pattern (\textbf{g}). 
    This nontrivial pattern results from the discrete three-fold rotation ($C_{3z}$) and mirror reflections ($M_y$ and $M_z$). 
    \textbf{d}, ZnSe (SG 186; type-III PG $C_{6v}$) combines six-fold rotation $C_{6z}$ and mirror symmetries ($M_x$ and $M_y$) to produce an $i$-wave helicity pattern (\textbf{h}), featuring a 12-fold alternation of sign within the Brillouin zone. 
    The energy of each surface in \textbf{e}–\textbf{h} corresponds to the purple dashed line in \textbf{a}–\textbf{d}, respectively.
    In \textbf{a}–\textbf{d}, helicity projection is represented by red (positive) and blue (negative) dots, with dot size proportional to the helicity amplitude. 
    The reference point $C$ in each isoenergy surface
    is marked with a green star in \textbf{e}–\textbf{h} and labelled by a green vertical line in \textbf{a}–\textbf{d}. The shaded planes in \textbf{f}–\textbf{h} denote the mirror symmetry planes in momentum space.}
    \label{fig:fig4}
\end{figure*}

To exemplify the diverse chiral phonon behaviors revealed through the high-throughput screening, we showcase four prototypical materials from CPMDB representing distinct classes of phonon helicity patterns, including \ce{LiAlO2} [\CPMDweb{mp-3427}, SG 92 ($P4_12_12$)] with $s$-wave helicity, \ce{AgClO4} [\CPMDweb{mp-22993}, SG 121 ($I\overline{4}2m$)] with $d$-wave helicity, \ce{NaSrP} [\CPMDweb{mp-13275}, SG 189 ($P\overline{6}2m$)] with $g$-wave helicity, and \ce{ZnSe} [\CPMDweb{mp-380}, SG 186 ($P6_3mc$)] with $i$-wave helicity. The phonon band structures for these selected materials, with bands colored by phonon helicity, are shown in Fig. \ref{fig:fig4}a-d.
For each prototype material, we calculated its isoenergy surfaces and helicity distributions near the $\Gamma$ point (Fig. \ref{fig:fig4}e-h) within a $\pm 0.2$-meV window centered on the energy level indicated in the band structures. These isoenergy surfaces provide a three-dimensional visualization of the helicity patterns and confirm the theoretical predictions outlined in Fig. \ref{fig:helicity}.
In \ce{LiAlO2}, the isoenergy surface displays uniform helicity coloring under the PG $D_4$, consistent with the $s$-wave pattern. In contrast, \ce{AgClO4} shows four distinct regions of alternating helicity under the PG $D_{2h}$, confirming its $d$-wave character. \ce{NaSrP} exhibits a six-fold alternating pattern characteristic of $g$-wave helicity under PG $D_{3h}$, while \ce{ZnSe} displays the most complex helicity texture with multiple sign changes consistent with the $i$-wave behavior under PG $C_{6v}$.
 
These four prototypes span part of the symmetry-enforced helicity patterns and demonstrate how PG constraints translate into distinct chiral phonon textures.
These quantitative helicity amplitudes and anisotropies offer concrete targets for polarized Raman, neutron, and ultrafast optical pump–probe experiments. 

\section{Conclusions and Discussions}

This comprehensive catalogue offers a robust statistical basis for identifying chiral phonon materials, fundamentally advancing our understanding of their prevalence and characteristics. By combining symmetry-based classification with a large-scale survey of \nbrofmats\ crystals, our work provides a quantitative statistical dataset of materials capable of hosting chiral phonons. This analysis shows that symmetry-allowed chiral phonon phenomena occur in approximately 24\% of the surveyed compounds—including a notable fraction in non-chiral SGs that have not been widely explored in prior studies. In particular, we distinguish between 7.5\% that support type-II ($s$-wave) chirality and 16.1\% that exhibit type-III behavior, characterized by intricate $d$-, $g$- or $i$-wave helicity patterns. These findings yield a list of 170 promising candidate materials spanning a wide range of chemistries, from insulators to metals, thereby greatly broadening the landscape for experimental investigation.
Our symmetry-derived classification scheme allows for predictive identification of chiral phonon candidates without reliance on computationally expensive phonon calculations, transforming the search for new materials from a largely empirical process to a systematic, principle-guided enterprise. 
Although we focus here on inorganic crystals with well-defined static symmetries, the symmetry-based classification is, in principle, applicable to hybrid and organic crystals \cite{Pols2024acs.jpclett.4c01629,Pols2025acs.nanolett.5c01708} once an appropriate PG is defined. Extending CPMDB to soft, dynamically disordered lattices will require going beyond harmonic phonons to capture temperature-dependent symmetry, strong anharmonicity, and disorder.
Moreover, our symmetry-based classification theory, constructed from magnetic PGs, extends naturally to TRS-breaking systems. TRS breaking—whether induced by magnetic order or coupling to external fields—can lower the crystal symmetry and lift band degeneracies, enabling non-zero phonon angular momentum and odd-parity helicity patterns (for example, $p$, $f$ or $h$ wave) even in centrosymmetric crystals.
In such cases, the symmetry allowance of phonon angular momentum should be evaluated using the reduced magnetic PG at the corresponding momentum. We have identified \nbrofmagnet\ magnetic crystals from our database for future characterization of their chiral phonon properties (Supplementary Information \cite{SI}, section B).

Despite the breadth of this catalogue, experimental validation remains paramount. Future research should prioritize direct detection of phonon angular momentum, leveraging state-of-the-art techniques such as time-resolved Raman spectroscopy, polarization-resolved neutron scattering, and ultrafast phonon interferometry. These experiments will be vital both for confirming our predictions and for establishing quantitative metrics for phonon circular dichroism in real materials.
For these non-thermally driven scenarios, it is particularly important to understand the electron-phonon coupling pathways. It is obvious that the selection rule depends on the symmetry of the electronic state, the elements of the electron phonon matrix, and the polarization of light. At the same time, the coupling between electrons and chiral phonons can also lead to the mutual coupling between phonon angular momentum and electron spin\cite{PhysRevB.108.064510}. These rich physical phenomena have guiding significance for selectively exciting and detecting chiral phonon modes experiments.

Furthermore, achieving controllable chirality in crystalline solids stands as a major challenge: crystals with fixed handedness are hindered by significant energetic barriers separating enantiomeric forms. Intriguingly, the higher-order phonon helicity materials highlighted here, while achiral in their equilibrium lattice states, can support transient and switchable chirality when driven by optical excitation or other external stimuli\cite{zeng2025photo}. This dynamic control over chiral phonon modes opens up new opportunities for the development of next-generation chiral catalysts and spintronic devices through spin-phonon interactions, unlocking functions inaccessible to static chiral lattices.

\textbf{Methods}
\textbf{Density functional theory calculations}
The Density Functional Theory calculations were performed with the {\it Vienna Ab-initio Simulation Package} \cite{vasp1,PhysRevB.48.13115} and the finite difference method. Interactions between ion cores and valence electrons were treated with the projector augmented-wave method~\cite{paw1}. The exchange-correlation effects were described using the generalized gradient approximation with the Perdew-Burke-Ernzerhof functional~\cite{PhysRevLett.77.3865}. Each calculated crystal structure is first fully relaxed, followed by construction of the dynamical matrix from DFT-derived force constant matrices via the post-processing software Phonopy~\cite{phonopy-phono3py-JPCM,phonopy-phono3py-JPSJ} and computation of the phonon band structure along high-symmetry paths and density of states. Specifically, for a given crystalline material, its force constant matrix is defined as the second-order derivative of the total energy with respect to the atomic displacements \cite{RevModPhys.73.515},
\begin{equation}
    \mathcal{F}_{\alpha\mu,\beta\nu}({\bf R}-{\bf R'})=\frac{\partial^2E}{\partial u_{\alpha\mu}({\bf R})\partial u_{\beta\nu}({\bf R'})} \mid_{u=0},
    \label{eq:fcmat}
\end{equation}
where {\bf R} and {\bf R}$^\prime$ represent translations of the Bravais lattice and $u_{\alpha\mu}(\bf R)$ ($u_{\beta\nu}(\bf R')$) refers to the atomic displacement of the atom labelled as $\mu$ ($\nu$) in the unit cell {\bf R} ({\bf R}$^\prime$) along the $\alpha$($\beta$) direction, where $\alpha,\beta=x,y,z$. The dynamical matrix is the mass-reduced discrete Fourier transform of the force constant matrix,

\begin{equation}
    \mathcal{D}_{\alpha\mu,\beta\nu}({\bf q})=\frac{1}{N_R\sqrt{M_{\mu}M_{\nu}}}\sum_{\bf R} e^{i \bf q\cdot R} \mathcal{F}_{\alpha\mu,\beta\nu}({\bf R}),
    \label{eq:dmat}
\end{equation}
where $M_{\mu}$ is the mass of atom $\mu$ and whose basis is defined as
\begin{equation}
    \begin{aligned}
        \ket{u_{\alpha\mu}({\bf q})}
        &= \frac{1}{\sqrt{N_R}}\sum_{\bf R}e^{-i{\bf q \cdot {\bf R}}} \ket{u_{\alpha\mu}({\bf R})} \\
        &= \frac{1}{\sqrt{N_R}}\sum_{\bf R}e^{-i{\bf q \cdot {\bf R}}} \ket{u_{\alpha}({\bf R}+\tau_{\mu})},
    \end{aligned}
    \label{eq:blochbasis}
\end{equation}
and $\tau_{\mu}$ is the position of atom $\mu$ in the unit cell ${\bf R}$.

For a given ${\bf q}$, the eigenvalues of the dynamical matrix $\mathcal{D}({\bf q})$ give the squares of phonon frequencies $\omega^2_n({\bf q})$ ($n=1,2,...,3N$ with $N$ being the number of atoms in one unit cell),
\begin{equation}
    \mathcal{D}({\bf q}) \ket{\phi_n^{\bf q}}= \omega^2_n({\bf q}) \ket{\phi_n^{\bf q}}.
    \label{eq:phonon}
\end{equation}
 Since the dynamical matrix $\mathcal{D}({\bf q})$ is Hermitian, the eigenvalues $\{\omega^{2}_n({\bf q})\}$ are real. The eigenvector $\ket{\phi_n^{\bf q}}=\Sigma_{\alpha\mu}C_n^{\alpha\mu}\ket{u_{\alpha\mu}({\bf q})}$ characterizes the lattice vibrational mode of frequency $\omega_n({\bf q})$ at $\bf q$.

In polar semiconductors, whose positive and negative ions lie on separate planes, the long range macroscopic electric field induced by long wavelength (with $\bf q \rightarrow 0$) longitudinal optical (LO) phonons is non-negligible \cite{RevModPhys.73.515}. The coupling between LO phonons and electric fields result in the difference in frequency between LO and transverse optical (TO) phonons at the Brillouin zone centre, namely the LO-TO splitting. 

To quantitatively characterize the LO-TO splitting of phonon frequencies, we need an additional non-analytical term correction (NAC) in the dynamical matrix. The origin of this term is the long range dipole-dipole interaction  given by \cite{gonze1997dynamical,wang2010mixed},

\begingroup
\thinmuskip = 1mu
\medmuskip = 1mu
\thickmuskip = 1mu
\begin{equation}
    \mathcal{D}^{NA}_{\alpha\mu,\beta\nu}({\bf q \rightarrow 0})= \frac{4\pi e^2}{\Omega\sqrt{M_{\mu}M_{\nu}}} \frac{(\bf q \cdot Z^{*}_{\mu})_{\alpha}(\bf q \cdot Z^{*}_{\nu})_{\beta}}{\bf q \cdot \epsilon_{\infty} \cdot q},
    \label{eq:NAC}
\end{equation} \endgroup
where $\Omega$ is the volume of one unit cell, and $\epsilon_{\infty}$ is the high frequency static dielectric tensor.  $Z^{*}_{\mu}$ is the Born effective charge tensor of atom $\mu$ in one unit cell, which describes the polarization ($\Vec{\mathcal{P}}$) induced by the displacement of atom $\mu$ ($\Vec{\mathcal{\tau}}_{\mu}$) under the condition of zero electric field ($\Vec{E}=0$) and defined as,
\begin{equation}
    Z^{*}_{\mu,\alpha\beta} = \Omega \frac{\partial\mathcal{P}_{\beta}}{\partial\tau_{\mu,\alpha}}\mid_{\Vec{E}=0},
    \label{eq:BORN_tensor}
\end{equation}
where $\alpha,\beta=x,y,z$.

Hence, the total dynamical matrix for $\bf q \rightarrow 0$ is expressed as,
\begin{equation}
    \mathcal{D}^{Total}({\bf q \rightarrow 0}) = \mathcal{D}({\bf q}) + \mathcal{D}^{NA}(\bf q),
\end{equation}
where $\mathcal{D}(\bf q)$ is the analytical term in equation (\ref{eq:dmat}). Note that the NAC term is 0 at $\bf q=0$.

Note that in the high-throughput calculations, small imaginary frequencies do appear in some materials. In the vast majority of cases, however, these imaginary values are extremely small and can be safely attributed to numerical errors rather than genuine structural instabilities.  Supplementary Information \cite{SI}, section E provides a detailed discussion.

\textbf{Phonon angular momentum and magnetic moments} Using the phonon eigenvectors and frequencies derived from precomputed phonon dispersion relation calculations, we systematically evaluate the phonon angular momentum and phonon magnetic moment for every mode.
In general, the (real) Bloch basis defined in equation (\ref{eq:blochbasis}) can be converted to a complex basis by the following unitary transformation:
\begin{align}
    & \ket{R_{\gamma\mu}({\bf q})} = \frac{\epsilon_{\alpha\beta\gamma}}{\sqrt{2}}(\ket{u_{\alpha\mu}({\bf q})}+i\ket{u_{\beta\mu}({\bf q})}), \\
    & \ket{L_{\gamma\mu}({\bf q})} = \frac{\epsilon_{\alpha\beta\gamma}}{\sqrt{2}}(\ket{u_{\alpha\mu}({\bf q})}-i\ket{u_{\beta\mu}({\bf q})}),
\end{align}
where $\epsilon_{\alpha\beta\gamma}$ is the Levi-Civita symbol and $\alpha,\beta,\gamma \in \{x,y,z\}$. The states $\ket{R_{\gamma\mu}({\bf q})}$ and $\ket{L_{\gamma\mu}({\bf q})}$ describe, respectively, right‑handed and left‑handed circularly‑polarized vibration modes of atom $\mu$ about the 
$\gamma$ axis. 

The phonon angular momentum operator $\hat{\textbf{J}}^{ph}=(\hat{J}_x^{ph},\hat{J}_y^{ph},\hat{J}_z^{ph})$ in wave vector $\mathbf{q}$, summed over all atoms in a unit cell, is then defined as \cite{zhang2014angular}
\begingroup
\thinmuskip = 0.5mu
\medmuskip = 0.5mu
\thickmuskip = 0.5mu
\begin{equation}
\scalebox{0.9}[1]{$
\displaystyle
    \hat{J}^{ph}_\gamma(\mathbf{q}) = \Sigma_{\mu} \ket{R_{\gamma\mu}({\bf q})}\bra{R_{\gamma\mu}({\bf q})} - \ket{L_{\gamma\mu}({\bf q})}\bra{L_{\gamma\mu}({\bf q})}.
    \label{eq:angmomentopt}$}
\end{equation} \endgroup
For a phonon normal mode $\ket{\phi_n^{\mathbf{q}}}$ (equation (\ref{eq:phonon})), the expectation value of its angular momentum is
\begin{equation}
    J^{ph}_\gamma(\mathbf{q}) = \bra{\phi_n^{\mathbf{q}}}\hat{J}^{ph}_\gamma(\mathbf{q})\ket{\phi_n^{\mathbf{q}}}.
\end{equation}
Throughout this work we set the reduced Planck constant to unity (that is, $\hbar\equiv1$).

In ionic materials, the phonon angular momentum will induce magnetic moments through the gyromagnetic ratio. The orbital magnetic moment of phonon for phonon mode $n$ in the wave vector $\mathbf{q}$ is given by~\cite{PhysRevMaterials.3.064405}:
\begin{equation}
\mathbf{M}_n(\mathbf{q})=\sum_{\mu}\mathbf{\gamma}_{\mu}\mathbf{J}_{n\mu}^{ph}(\mathbf{q}),
\end{equation}
where $\mathbf{\gamma}_{\mu}=e\mathbf{Z}_{\mu}^*/(2M_{\mu})$ are the gyromagnetic ratio tensor of the ions $\mu$, $\mathbf{Z}_{\mu}^*$ are the Born effective charge tensors defined in equation (\ref{eq:BORN_tensor}), $M_{\mu}$ are the masses and $e$ denotes the elementary charge; here we neglect the off-diagonal terms.   $\mathbf{J}_{n\mu}^{ph}(\mathbf{q})=(J^{ph}_{n\mu x}(\mathbf{q}),J^{ph}_{n\mu y}(\mathbf{q}),J^{ph}_{n\mu z}(\mathbf{q}))$ are the angular moments which means $J^{ph}_{n\mu \gamma}(\mathbf{q})=\bra{\phi_{n\mu}^{\mathbf{q}}}\hat{J}^{ph}_{\mu \gamma}(\mathbf{q})\ket{\phi_{n\mu}^{\mathbf{q}}}$ with:
\begingroup
\thinmuskip = 1mu
\medmuskip = 1mu
\thickmuskip = 1mu
\begin{equation}
\scalebox{0.95}[1]{$
\displaystyle
    \hat{J}^{ph}_{\mu \gamma}(\mathbf{q})=\ket{R_{\gamma\mu}({\bf q})}\bra{R_{\gamma\mu}({\bf q})} - \ket{L_{\gamma\mu}({\bf q})}\bra{L_{\gamma\mu}({\bf q})}$}
\end{equation} \endgroup
is the operator of phonon circular polarization of atom $\mu$ in direction $\gamma$. And $\ket{\phi_{n\mu}^{\mathbf{q}}}$ is the unit eigenvector of the ion $\mu$~\cite{PhysRevB.105.024312}. Supplementary Information \cite{SI}, section F shows the calculations of phonon magnetic moments in real materials.

\textbf{Phonon helicity and cycloidicity} 
Not all phonons carrying angular momentum are chiral, as circular motion may or may not break symmetry. We calculated the helicity $h$ and cycloidicity vector \textbf{P} to characterize the chirality of phonons. In detail, for an arbitrary lattice momentum $\mathbf{q}$ in a non-centrosymmetric three-dimensional Brillouin zone, its corresponding little group $G_{\mathbf{q}}$ is typically trivial, $G_{\mathbf{q}}=\{E\}$, that is, it contains only the identity operation. Consequently, the phonon angular momentum at this momentum $\mathbf{J}(\mathbf{q})$ is generically nonzero. Moreover, the relative orientation between $\mathbf{J}(\mathbf{q})$ and $\mathbf{q}$ is, in general, arbitrary. One may therefore decompose
\[
\begin{aligned}
\mathbf{J}(\mathbf{q})= \mathbf{J}_{\parallel}(\mathbf{q})
 + \mathbf{J}_{\perp}(\mathbf{q}),\\
\mathbf{J}_{\parallel}(\mathbf{q})= \big(\mathbf{e}_q\cdot\mathbf{J}(\mathbf{q})\big)
   \mathbf{e}_q,\\
\mathbf{J}_{\perp}(\mathbf{q})= \mathbf{J}(\mathbf{q})
 - \mathbf{J}_{\parallel}(\mathbf{q}),
\end{aligned}
\]
where $\mathbf{e}_q=\mathbf{q}/|\mathbf{q}|$.
The component parallel to $\mathbf{q}$ defines the \emph{helical} part and is quantified by the pseudoscalar helicity \cite{ishito2023truly,ishito2023chiral}, $h(\mathbf{q})=\mathbf{e}_q\cdot \mathbf{J}(\mathbf{q})$.
A non-zero component of $h(\mathbf{q})$ dynamically breaks improper rotational symmetry. The component perpendicular to $\mathbf{q}$ defines the \emph{cycloidal} part, whose symmetry properties can be characterized by the polar vector, $\mathbf{P}(\mathbf{q})=\mathbf{e}_q\times \mathbf{J}(\mathbf{q})$.

When a component of \textbf{P} is non-zero, the circular polarization in combination with a propagation of the displacement along this direction does not break improper rotational symmetry. Unlike the lattice momentum $\mathbf{q}$, $\mathbf{P}$ is time-reversal invariant. Under point-group operations, $\mathbf{P}$ transforms according to the same representation as $\mathbf{q}$. Accordingly, the $\mathbf{P}$ texture near $\Gamma$ ($\mathbf{q}=0$) in the Brillouin zone transforms as a polar vector under the $\Gamma$-point symmetry group. This representation decomposition has already been carried out for $\mathbf{q}$; for completeness, we list the corresponding decomposition for $\mathbf{P}$ again in Supplementary Information \cite{SI}, section A.

\textbf{Pseudo-angular momentum}
To characterize the chiral signature of phonon modes, we calculate the phononic pseudo-angular momentum (PAM) \cite{zhang2015chiral} for all material entries based on the symmetry eigenvalues of phonon modes. PAM decomposes into (1) a local spin contribution and (2) a non‑local orbital contribution, capturing, respectively, the intracell and intercell responses to (screw) rotations \cite{zhang2015chiral,PhysRevB.105.104301,PhysRevResearch.4.L012024,zhang2026chirality}. To make it concrete, we derive the explicit PAM expressions below following the conventions of ref. \cite{xu2024catalog}.

Given a space group $\mathcal{G}$, a general symmetry operator $g \in \mathcal{G}$ is defined as $g = \{\mathcal{R}|{\tau}\}$ where $\mathcal{R}$ is a symmetry operation of the point group and $\tau$ represents a translation. For symmorphic space groups expressed in a primitive basis and with a proper choice of the origin, $\tau$ has integer components in all symmetry operations. However, in non-symmorphic space groups, the translation $\tau$ of certain symmetry operations takes fractional values. In reciprocal space, the little group of a {\bf q}-point is $G_{\bf q} \subset \mathcal{G}$, whose elements $\tilde{g}=\{\mathcal{\tilde{R}}|\tilde{\tau}\}$ satisfy the following condition, 
\begin{equation}
   \mathcal{\tilde{R}} {\bf q} \cdot = {\bf q} +{\bf G}\label{eq:lg}
\end{equation}  
where ${\bf G}$ are the reciprocal lattice vectors.

In real materials, the dynamical matrix is written in the basis of $\ket{u_{\alpha\mu}({\bf q})}$. When a general symmetry operator $\tilde{g}=\{\mathcal{\tilde{R}}\mid\tilde{\tau}\}$ acts on $\ket{u_{\alpha\mu}({\bf q})}$, we have,

{\small
\begingroup
\thinmuskip = 0.2mu
\medmuskip = 0.2mu
\thickmuskip = 0.2mu

\begin{equation}
\scalebox{0.9}[1]{$
\displaystyle
    \begin{aligned}
    \{\mathcal{\tilde{R}}|\tilde{\tau}\} \ket{u_{\alpha\mu}({\bf q})}
    &=\frac{1}{\sqrt{N_R}}\sum_R e^{-i{\bf q}\cdot {\bf R}} \ket{u_{\tilde{\mathcal{R}}\alpha}(\mathcal{\tilde{R}}{\bf R}+\tilde{\mathcal{R}}\tau_{\mu}+\tilde{\tau})} \\
    &=\frac{1}{\sqrt{N_R}}\sum_R e^{-i(\tilde{\mathcal{R}}{\bf q})\cdot \tilde{\mathcal{R}}{\bf R}} \ket{u_{\tilde{\mathcal{R}}\alpha}(\mathcal{\tilde{R}}{\bf R}+\tilde{\mathcal{R}}\tau_{\mu}+\tilde{\tau})} \\
    &=\frac{1}{\sqrt{N_{R^\prime}}}\sum_{R^\prime,\mu\prime} e^{-i{\bf q} [{\bf R^\prime}+\tau_{\mu^\prime}-\tilde{\mathcal{R}}\tau_{\mu}-\tilde{\tau}]} \ket{u_{\tilde{\mathcal{R}}\alpha}({\bf R^\prime}+\tau_{\mu^\prime})} \\
    &=\sum_{\mu\prime,\alpha\prime}\mathcal{O}_{\alpha^{\prime}\alpha}e^{-i{\bf q}[\tau_{\mu^\prime}-\tilde{\mathcal{R}}\tau_{\mu}-\tilde{\tau}]}  \frac{1}{\sqrt{N_{R^\prime}}} \sum_{R^\prime}e^{-i{\bf q}{\bf R^\prime}} \\ &~~~\ket{u_{\alpha\prime}({\bf R^\prime}+\tau_{\mu\prime})} \\
    &= \sum_{\mu^\prime,\alpha^\prime} \mathcal{O}_{\alpha^{\prime}\alpha} \mathcal{A}_{\mu^{\prime}\mu} \ket{u_{\alpha^{\prime}\mu^{\prime}}({\bf q})},
    \end{aligned}
    $}
    \label{eq:symmrep}
\end{equation} \endgroup}
where we have taken $\mathcal{\tilde{R}}{\bf R}+\tilde{\mathcal{R}}\tau_{\mu}+\tilde{\tau}={\bf R^\prime}+\tau_{\mu\prime}$, and $\ket{u_{\alpha^{\prime}\mu^{\prime}}({\bf q})}=\frac{1}{\sqrt{N_{R^\prime}}}\sum_{R^\prime}e^{-i{\bf q}{\bf R^\prime}} \ket{u_{\alpha\prime}({\bf R^\prime}+\tau_{\mu\prime})}$. $\mathcal{O}$ and $\mathcal{A}$ are the transformation matrices of the vibrational directions and atomic positions under the operation $\tilde{g}$, respectively. 
As derived above, the element of $\mathcal{A}$ is expressed as
{\small
\begingroup

\thinmuskip = 0.2mu
\medmuskip = 0.2mu
\thickmuskip = 0.2mu
\begin{equation}
    \mathcal{A}_{\mu^\prime\mu}=
    \left\{
    \begin{array}{@{}l@{\;}l@{}}
         e^{-i {\bf q}\cdot [\tau_{\mu\prime}-(\tilde{\mathcal{R}}\tau_{\mu}+\tilde{\tau})]},&\text{when~} \tau_{\mu\prime}=(\tilde{\mathcal{R}}\tau_{\mu}+\tilde{\tau}) \text{~mod~} {\bf R}  \\
         0, & \text{otherwise.} \\
    \end{array}
    \right.
    \label{eq:A}
\end{equation}
\endgroup
}

The matrix $\mathcal{O}$ is actually the representation of $\mathcal{\tilde{R}}$ under the basis of Cartesian coordinates, \ie $[\mathcal{\tilde{R}}]_{c}$. However, in the standard setting of the space group $\mathcal{G}$, the point group operations $\mathcal{\tilde{R}}$ are expressed in the basis of the lattice of translations given by three vectors ${\bf a_{i=1,2,3}}$ which are, in general, not orthogonal. If we represent the symmetry operation in the basis of the lattice as $[\mathcal{\tilde{R}}]_{\ell}$, the transformation between the two matrices is given by,
\begin{equation}
    \mathcal{O}=[\mathcal{\tilde{R}}]_{c} =  U[\mathcal{\tilde{R}}]_{\ell}U^{-1}
    \label{eq:O}
\end{equation}
where the transformation matrix is
\begin{equation}
    U=    \left[
    \begin{array}{l}
         {\bf a_1} \\
         {\bf a_2} \\
         {\bf a_3} \\
    \end{array}
    \right]^T,
    \label{eq:U}
\end{equation}
and ${\bf a_{i=1,2,3}}$ are the three lattice vectors expressed in Cartesian coordinates. 

In three-dimensional crystals, the only admissible discrete rotational symmetries are $C_n$ ($n=2,3,4,6$), whose eigenvalues take the form $e^{-i\frac{2\pi}{n}l}$  ($l = 0,1, \cdots, n-1$).
Consider a phonon‐mode eigenstate $\lvert \phi_{n}^{\mathbf q}\rangle$ whose little group at wavevector $\mathbf q$ contains the (screw) rotation $\{C_n\mid\tau\}$.  The expectation value of its symmetry operator,
\begin{equation}
    e^{-i\frac{2\pi}{n}l^{ph}}=\langle \phi_{n}^{\mathbf q} \rvert\mathcal{O}\otimes\mathcal{A}\lvert \phi_{n}^{\mathbf q}\rangle,
\end{equation}
yields the phase factor $e^{-2\pi i\,l^{\rm ph}/n}$, where $\mathcal{O}\otimes\mathcal{A}$ is the matrix representation of $\{C_n\mid\tau\}$ (equations (\ref{eq:symmrep})–(\ref{eq:U})) and $l^{\rm ph}$ is the PAM of the mode.

\acknowledgments
This work was supported by the National Key R\&D Program of China (grant numbers 2025YFA1411500 to Y.X., Y.L. and M.S. and 2022YFA1402200), the Zhejiang Provincial Natural Science Foundation of China (grant numbers LR26A040003 to Y.X., LR24A040001 to Y.T. and LJXSZ26A040002 to T.D.), the National Natural Science Foundation of China (grant numbers 12550401 to Y.X., 12374163 to Y.X., W2511006 to H.Y., 12494592 to H.Y., 12350710785 to M.S., 12561160109 to M.S., 62574179 to T.D. and 12274365 to Y.T.), the New Cornerstone Science Foundation to H.Y., and the Fundamental Research Funds for the Central Universities (grant numbers 226-2024-00200 to Y.X. and 226-2024-00068 to M.S.). Part of the calculations was performed on Quantum Many-Body HPC platform at School of Physics, Zhejiang University.

\vspace{1cm}
{\bf Author Contributions}
Y.X. conceived and supervised the project. Y.Y., Z.X., Z.L. and Y.X. developed the classification method. Y.Y. and Y.M. performed the high-throughput calculations under the supervision of Y.X. Z.W. built the database using input data provided by Y.Y. and Y.M.. T.D., Y.T., Z.S., Y.L., H.Y., M.S., and Y.X. analyzed and discussed the results. Y.X., Y.Y.,  Z.X. and Y.M. drafted the manuscript and Supplementary Information \cite{SI}, and all authors contributed to revising the final version.

{\bf Competing Interests}
The authors declare no competing interests.

{\bf Data Availability}
Source data are provided with this paper. All other data are publicly 
available in the CPMDB at \url{https://materialsfingerprint.com}.

{\bf Code Availability}
The code used in this study is available via Code Ocean at \url{https://doi.org/10.24433/CO.0627690.v1}.

\bibliography{chirality}
\end{document}

%% file: macros.tex
\newcommand{\CPMDweb}[1]{\href{https://materialsfingerprint.com/result/#1}{#1}}

\newcommand{\nbrofmats}{10,219}
\newcommand{\nbrofmpid}{9,989}
\newcommand{\nbroficsd}{1,625}

\newcommand{\nbroftypeI}{7,770}
\newcommand{\nbroftypeIpercent}{76.03\%}

\newcommand{\nbroftypeII}{808}
\newcommand{\nbroftypeIIpercent}{7.91\%}

\newcommand{\nbroftypeIII}{1,641}
\newcommand{\nbroftypeIIIpercent}{16.06\%}

\newcommand{\nbrofdwave}{1,011}
\newcommand{\nbrofdwavepercent}{9.89\%}
\newcommand{\nbrofgwave}{286}
\newcommand{\nbrofgwavepercent}{2.80\%}
\newcommand{\nbrofiwave}{344}
\newcommand{\nbrofiwavepercent}{3.37\%}

\newcommand{\nbrofideal}{170}

\newcommand{\nbrofmatstmdb}{31,937}

\newcommand{\nbrofmagnet}{241}
\newcommand{\nbrofmatsmagmom}{494}